\documentclass{ejm}

\begin{document}

%my macros for LaTeX fixes

% refs: 
%put `( )' around an equation ref in latex
\def\eqref#1{(\ref{#1})}
\def\eqrefs#1#2{(\ref{#1}) and~(\ref{#2})}
\def\eqsref#1#2{(\ref{#1}) to~(\ref{#2})}
\def\sysref#1#2{(\ref{#1}),~(\ref{#2})}

%put `Eq.( )' around an equation ref in latex
\def\Eqref#1{Eq.~(\ref{#1})}
\def\Eqrefs#1#2{Eqs.~(\ref{#1}) and~(\ref{#2})}
\def\Eqsref#1#2{Eqs.~(\ref{#1}) to~(\ref{#2})}
\def\Sysref#1#2{Eqs. (\ref{#1}),~(\ref{#2})}

%put `Sec. ' before a section ref in latex
\def\secref#1{Sec.~\ref{#1}}
\def\secrefs#1#2{Sec.~\ref{#1} and~\ref{#2}}

%put `App. ' before an appendix ref in latex
\def\appref#1{Appendix~\ref{#1}}

%put `Ref. ' before a bibitem ref in latex
\def\Ref#1{Ref.~\cite{#1}}

%use footnote style for a bibitem ref in latex
\def\Cite#1{${\mathstrut}^{\cite{#1}}$}

%put `Table ' before a table ref in latex
\def\tableref#1{Table~\ref{#1}}

%put `Fig. ' before a figure ref in latex
\def\figref#1{Fig.~\ref{#1}}

% fix hyphenations to be 
\hyphenation{Eq Eqs Sec App Ref Fig}

% abbrevs for latex commands:
%equations
\def\EQ{\begin{equation}}
\def\EQs{\begin{eqnarray}}
\def\endEQ{\end{equation}}
\def\endEQs{\end{eqnarray}}

%theorems, definitions, proofs
\def\Thm{\begin{theorem}}
\def\endThm{\end{theorem}}
\def\Lem{\begin{lemma}}
\def\endLem{\end{lemma}}
\def\Cor{\begin{corollary}}
\def\endCor{\end{corollary}}
\def\Prop{\begin{proposition}}
\def\endProp{\end{proposition}}
\def\Defn{\begin{definition}}
\def\endDefn{\end{definition}}
\def\Proof[#1]{\begin{proof}[#1]}
\def\endProof{\end{proof}}

\def\Rem[#1]{{\bf {#1}}}
\def\proclaim#1{\medbreak
\noindent{\it {#1}}\par\medbreak}

%my macros

\def\fewquad{\qquad\qquad}
\def\severalquad{\qquad\fewquad}
\def\manyquad{\qquad\severalquad}
\def\manymanyquad{\manyquad\manyquad}

\def\downupindices#1#2{{}^{}_{#1}{}_{}^{#2}}
\def\updownindices#1#2{{}_{}^{#1}{}^{}_{#2}}
\def\mixedindices#1#2{{\mathstrut}^{#1}_{\mathstrut #2}}
\def\downindex#1{{\mathstrut}^{\mathstrut}_{#1}}
\def\upindex#1{{\mathstrut}_{\mathstrut}^{#1}}

\def\eqtext#1{\hbox{\rm{#1}}}

\def\Parder#1#2{
\mathchoice{\partial{#1} \over\partial{#2}}{\partial{#1}/\partial{#2}}{}{} }
\def\nParder#1#2#3{
\mathchoice{\partial^{#1}{#2} \over\partial{#3}^{#1}}{\partial^{#1}{#2}/\partial{#3}^{#1}}{}{} }
\def\mixedParder#1#2#3#4{
\mathchoice{\partial^{#1}{#2} \over\partial{#3}\partial{#4}}{\partial^{#1}{#2}/\partial{#3}\partial{#4}}{}{} }
\def\parder#1#2{\partial{#1}/\partial{#2}}
\def\nparder#1#2{\partial^{#1}/(\partial {#2})^{#1}}
\def\mixedparder#1#2#3#4{\partial^{#1}{#2}/\partial{#3}\partial{#4}}
\def\parderop#1{\partial/\partial{#1}}
\def\mixedparderop#1#2#3{\partial^{#1}/\partial{#2}\partial{#3}}

\def\x#1{x^{#1}}
\def\u#1{u\upindex{#1}}
\def\bdx{{\boldsymbol x}}
\def\bdu{{\boldsymbol u}}
\def\ujet#1{\nder{}{#1}u}
\def\bdujet#1{\nder{}{#1}\bdu}
\def\uder#1#2{u\mixedindices{#1}{#2}}
\def\deru#1#2{\der{#2}{\u{#1}}}
\def\nderu#1#2#3{\nder{#2}{#3}\u{#1}}
\def\purederu#1#2#3{\nParder{#2}{\u{#1}}{#3}}
\def\mixderu#1#2#3#4{\mixedParder{#2}{\u{#1}}{#3}{#4}}

\def\D#1{D\downindex{#1}}
\def\G#1{G\upindex{#1}}
\def\linop#1#2#3{({\cal L}_{#1})\mixedindices{#2}{#3}}
\def\adlinop#1#2#3{({\cal L}^*_{#1})\mixedindices{#2}{#3}}
\def\symm#1{\eta\upindex{#1}}
\def\adsymm#1{\omega\downindex{#1}}
\def\factor#1{\Lambda\downindex{#1}}  
\def\bdfactor{{\boldsymbol \Lambda}}
\def\Udersymm#1#2#3{\Parder{\symm{#1}}{\uder{#2}{#3}}}
\def\Uderadsymm#1#2#3{\Parder{\adsymm{#1}}{\uder{#2}{#3}}}
\def\Uderfactor#1#2#3{\Parder{\factor{#1}}{\uder{#2}{#3}}}
\def\Dfactor#1#2{\Lambda\downindex{#1#2}}

\def\factorseq#1{\factor{(#1)}}

\def\ujetl{\ujet{\ell}}

\def\g#1{g\upindex{#1}} 
\def\bdg{{\boldsymbol g}}
\def\Uderg#1#2#3{\Parder{\g{#1}}{\uder{#2}{#3}}}
\def\parderg#1#2{\Parder{\g{#1}}{{#2}}}

\def\M#1{M\downindex{#1}}
\def\P#1{\Phi\upindex{#1}}
\def\S#1{S\upindex{#1}}
\def\Phat#1{{\tilde\Phi}\upindex{#1}}  
\def\K{K}
\def\Y#1{A\upindex{#1}}
\def\T#1{T\upindex{#1}}
\def\C#1#2{C\mixedindices{#2}{#1}}

\def\w#1{w\downindex{#1}}
\def\v#1{v\upindex{#1}}
\def\bdv{{\boldsymbol v}}
\def\W#1{W\downindex{#1}}
\def\V#1{V\upindex{#1}}
\def\Wt#1{W\upindex{#1}}
\def\vder#1#2{v\mixedindices{#1}{#2}}

\def\tU#1{{\tilde U}\upindex{#1}}
\def\tu#1{{\tilde u}\upindex{#1}}  
\def\bdtu{{\tilde \bdu}}  
\def\tuder#1#2{{\tilde u}\mixedindices{#1}{#2}}
\def\uparm#1#2{\uder{#1}{(#2)}}
\def\uparmder#1#2#3{\uder{#1}{(#2)#3}}
\def\ul{u\downindex{(\lambda)}}
\def\uljet#1{\nder{}{#1}u\downindex{(\lambda)}}
\def\bdul{\bdu\downindex{(\lambda)}}
\def\bduljet#1{\nder{}{#1}\bdu\downindex{(\lambda)}}

\def\Pder#1#2{\Phi\mixedindices{#1}{#2}}

\def\X#1{{\rm X}_{#1}}
\def\ELop#1#2{E_{#1\upindex{#2}}}
\def\parderU#1#2{\der{\uder{#1}{#2}}}
\def\ELophat#1#2{{\hat E}_{#1\upindex{#2}}} 
\def\elop#1{E_{#1}}
\def\elophat#1{{\hat E}_{#1}}

\def\nD#1#2{D\mixedindices{#1}{#2}}
\def\der#1{\partial_{#1}}
\def\nder#1#2{\partial^{#2}_{#1}}
\def\Dt#1{D\mixedindices{#1}{t}}
\def\Dx#1{D\mixedindices{#1}{x}}
\def\solD#1{{\cal D}\downindex{#1}}
\def\solnD#1#2{{\cal D}\mixedindices{#1}{#2}}

\def\Du#1{u_{#1}}
\def\bdDu#1{{\bdu}_{#1}}
\def\Dtu#1{{\tilde u}_{#1}}
\def\bdDtu#1{{\tilde \bdu}_{#1}}  
\def\bdDul#1{\bdu\downindex{(\lambda){#1}}}
\def\Dv#1{v_{#1}}

\def\slinop#1{{\cal L}_{#1}}
\def\adslinop#1{{\cal L}^*_{#1}}

\def\N{N}
\def\measure{{d\lambda \over\lambda}}
\def\tc{\tilde c}

\def\h{h}

\def\pde/{partial differential equation}
\def\de/{differential equation}
\def\conslaw/{conservation law}
\def\adsys/{adjoint system}
\def\adinv/{adjoint invariance condition}
\def\hsys/{extra system}
\def\dcm/{direct conservation law method}
\def\CK/{Cauchy-Kovalev\-skaya}

\def\ie/{i.e.}
\def\eg/{e.g.}
\def\etc/{etc.}
\def\const{{\rm const}}

%KdV macros

\def\pot{\psi}
\def\Dpot#1{\psi_{#1}}
\def\Pot{\psi}
\def\DPot#1{\psi_{#1}}
\def\potsymm{\varrho}
\def\potsymmseq#1{\varrho_{(#1)}}
\def\wder#1{\potsymm\downindex{#1}}

\def\linG{{\cal L}_G}
\def\adlinG{{\cal L}^*_G}
\def\linl{{\cal L}_\l}
\def\adlinl{{\cal L}^*_\l}
\def\lin#1{{\cal L}_{#1}}
\def\adlin#1{{\cal L}^*_{#1}}   

\def\invDx{D\mixedindices{-1}{x}}

\def\symmseq#1{\eta_{(#1)}}
\def\adsymmseq#1{\omega_{(#1)}} 

\def\l{\Lambda}
\def\lhat{{\hat\Lambda}}
\def\lder#1{\Lambda\downindex{#1}}
\def\adsymmder#1{\omega\downindex{#1}} 
\def\adsymmseqder#1#2{\omega_{(#1)#2}} 

\def\lparmu{\lambda u}

%KG macros

\def\Dhat#1{{\hat D}\downindex{#1}}

\newtheorem{theorem}{Theorem}[subsection]
\newtheorem{corollary}[theorem]{Corollary}

%\preprint{ To appear in Eur. J. Appl. Math }
% date of version: Aug 22, 2001

\title[Direct conservation law method]
{ Direct construction method for conservation \\
laws of partial differential equations. Part I: \\
Examples of conservation law classifications }

\author[ S.\ns C.\ns A\ls N\ls C\ls O\ns
\and G.\ns B\ls L\ls U\ls M\ls N\ls A\ls N]
{ S\ls T\ls E\ls P\ls H\ls E\ls N\ns C.\ns A\ls N\ls C\ls O$\,^1$\ns
\and G\ls E\ls O\ls R\ls G\ls E\ns B\ls L\ls U\ls M\ls A\ls N$\,^2$ }
\affiliation{
$^1\,$Department of Mathematics, 
Brock University,
St. Catharines, ON Canada L2S 3A1\\
email\textup{\nocorr: \texttt{sanco@brocku.ca}}\\
$^2\,$Department of Mathematics, 
University of British Columbia, 
Vancouver, BC Canada V6T 1Z2 \\
email\textup{\nocorr: \texttt{bluman@math.ubc.ca}} }

\maketitle

\begin{abstract}
An effective algorithmic method is presented 
for finding the local conservation laws 
for partial differential equations 
with any number of independent and dependent variables. 
The method does not require the use or existence of a variational principle
and reduces the calculation of conservation laws 
to solving a system of linear determining equations 
similar to that for finding symmetries. 
An explicit construction formula is derived which 
yields a conservation law for each solution of the determining system.  
In the first of two papers (Part I), 
examples of nonlinear wave equations
are used to exhibit the method. 
Classification results for conservation laws of these equations
are obtained.
In a second paper (Part II),
a general treatment of the method is given. 
\end{abstract}

%\newpage

\section{Introduction}
\label{intro}

In the study of differential equations, 
\conslaw/s have many significant uses, 
particularly with regard to  
integrability and linearization,
constants of motion, 
analysis of solutions,
and numerical solution methods. 
Consequently, an important problem is
how to calculate all of the \conslaw/s for given differential equations.

For a differential equation with a variational principle, 
Noether's theorem \cite{noether,noether2,noether3,blumanbook,olverbook}
gives a formula for obtaining the local \conslaw/s
by use of symmetries of the action. 
One usually attempts to find these symmetries by noting that 
any symmetry of the action leaves invariant the extremals of the action 
and hence gives rise to a symmetry of the differential equation. 
However, all symmetries of a differential equation 
do not necessarily arise from symmetries of the action
when there is a variational principle.
For example, if a differential equation is scaling invariant, 
then the action is often not invariant. 
Indeed it is often computationally awkward 
to determine the symmetries of the action 
and carry out the calculation with the formula
to obtain a \conslaw/. 
Moreover, in general 
a differential equation need not have a variational principle 
even allowing for a change of variables. 
Therefore, 
it is more effective to seek 
a direct, algorithmic method without involving an action principle 
to find the \conslaw/s of a given differential equation. 

In \Ref{prlpaper} 
we presented an algorithmic approach 
replacing Noether's theorem
so as to allow one to obtain all local \conslaw/s 
for any differential equation whether or not it has a variational principle. 
Details of this approach for the situation of 
ordinary differential equations (ODEs) 
are given in \Ref{odepaper}. 
Here we concentrate on the situation of partial differential equations (PDEs). 

In the case of a PDE with a variational principle, 
the approach shows how to use the symmetries of the PDE
to directly construct the \conslaw/s. 
The symmetries of a PDE satisfy 
a linear determining equation, 
for which there is a standard algorithmic method 
\cite{blumanbook,olverbook}
to seek all solutions. 
There is also an invariance condition, 
involving just the PDE and its symmetries,
which is necessary and sufficient 
for a symmetry of a PDE with a variational principle 
to correspond to a symmetry of the action. 
The invariance condition can be checked by an algorithmic calculation
and, in addition, leads to a direct construction formula
for a \conslaw/ in terms of the symmetry and the PDE. 
This approach makes no use of the variational principle for the PDE.

In the case of a PDE without a variational principle, 
the approach involves replacing symmetries by adjoint symmetries of the PDE.
The adjoint symmetries satisfy a linear determining equation that is
the adjoint of the determining equation for symmetries.
Geometrically, symmetries of a PDE describe motions 
on the solution space of the PDE.
Adjoint symmetries in general do not have such an interpretation.  
The invariance condition on symmetries is replaced by 
an adjoint invariance condition on adjoint symmetries
and there is a corresponding direct construction formula
for obtaining the \conslaw/s in terms of the adjoint symmetries and the PDE.
The adjoint invariance condition is a necessary and sufficient 
determining condition for an adjoint symmetry to yield a \conslaw/.

In general for any PDE, with or without a variational principle, 
the approach of \Ref{prlpaper} for finding all local \conslaw/s 
gave the following step-by-step method:
  
(1) Find the adjoint symmetries of the given PDE.

(2) Check the adjoint invariance condition 
on the adjoint symmetries.

(3) For each adjoint symmetry satisfying the adjoint invariance condition
use the direct construction formula to obtain a \conslaw/. 

If the adjoint symmetry determining equation is 
the same as the symmetry determining equation
then adjoint symmetries are symmetries. 
In this case the given PDE can be shown to have a variational principle. 
Conversely, if a variational principle exists, 
then the symmetry determining equation of the PDE 
can be shown to be self-adjoint,
so that symmetries are adjoint symmetries.  

In order to solve the determining equation for adjoint symmetries,
one works on the solution space of the given PDE. 
On the other hand, in order to check the adjoint invariance condition,
one must move off the solution space by replacing 
the dependent variable(s) of the given PDE by functions
with arbitrary dependence on the independent variables of the given PDE. 
The same situation arises when checking for invariance of the action
in Noether's theorem. 

These steps of the method are an algorithmic version of
the standard treatment presented in \Ref{olverbook} 
for finding PDE \conslaw/s in terms of multipliers. 
In particular, multipliers can be characterized as 
adjoint symmetries that satisfy the adjoint invariance condition
and thus can be calculated by the step-by-step algorithm (1), (2), (3). 

In this paper (Part I) and a sequel (Part II), 
we significantly improve 
the effectiveness of this method
by replacing the adjoint invariance condition by 
extra determining equations
which allow one to work entirely on the solution space of the given PDE.
Consequently, by augmenting the adjoint symmetry determining equation 
by these extra determining equations 
we obtain a linear determining system 
for finding only those adjoint symmetries that 
are multipliers yielding \conslaw/s. 
At first sight it is natural to proceed as in \Ref{prlpaper} by 
solving the adjoint symmetry determining equation
and then checking which of the solutions 
satisfy the extra determining equations.
On the other hand, 
all of the determining equations are on an equal footing,
and hence there is no requirement 
to solve the adjoint symmetry determining equation first.
Indeed, as illustrated later in the examples in this paper,
it is much more effective to start with the extra determining equations 
before considering the adjoint symmetry determining equation. 
This is true even in the case when the given PDE has a variational principle.

In solving the determining system 
one works completely on the solution space of the given PDE. 
Hence one can use the same algorithmic procedures 
as for solving symmetry determining equations 
in order to solve the \conslaw/ determining system.
In particular, existing symbolic manipulation programs 
\cite{software}
that calculate symmetries can be readily adapted to calculate
solutions of the \conslaw/ determining system. 
Moreover, for each solution one can directly obtain the resulting \conslaw/
by evaluating the construction formula 
working entirely on the solution space of the given PDE.

The \conslaw/ determining system together with 
the \conslaw/ construction formula 
give a general, direct, computational method 
for finding the local \conslaw/s of given PDEs. 
We refer to this as the direct \conslaw/ method. 
As emphasized above, 
its effectiveness stems from allowing
the calculation of \conslaw/s to be carried out algorithmically
up to any given order 
by solving a linear determining system
without moving off the solution space of the PDEs.
Compared to the standard treatment of PDE \conslaw/s, 
the determining system solves a long-standing question of
how one can delineate necessary and sufficient determining equations
to find multipliers 
by working entirely on the solution space of the given PDE. 
Most importantly, 
by mingling the adjoint symmetry equations with the extra equations
in the determining system, 
one can gain a significant computational advantage 
over the standard methods for finding multipliers. 

In \secref{examples} 
we illustrate the \dcm/ 
through classifying \conslaw/s for three PDE examples: 
a generalized Korteweg-de Vries equation, 
a nonlinear wave-speed equation,
and a class of nonlinear Klein-Gordon equations. 
The classification results obtained are new in that
they establish the completeness of certain families of \conslaw/s
which are of interest for these PDEs. 
These examples 
show how to calculate all \conslaw/s up to a given order
and also how to determine which PDEs in a specified class admit
\conslaw/s of a given type. 

In the second paper (Part II), 
we present a general derivation of 
the \conslaw/ determining system and construction formula,
and we also give a summary of the general method.

\section{Examples of conservation law classifications}
\label{examples}

Here we illustrate the use of our \dcm/
on three PDE examples.
For each example we derive the \conslaw/ determining system
and use it to obtain a classification result for \conslaw/s.

The first example is a 
generalized Korteweg-de Vries equation in physical form, 
for which there is no direct variational principle.
The ordinary Korteweg-de Vries (KdV) equation 
is well-known to have local \conslaw/s of every even order \cite{kdv},
which can be understood to arise 
from a recursion operator \cite{olver}.
Using our \conslaw/ determining system, 
we derive a direct, complete classification for
all \conslaw/s up to second order
for the generalized KdV equation. 
This example illustrates a general approach 
for finding and classifying \conslaw/s for 
non-variational evolution equations.

The second example is a scalar wave equation with non-constant wave speed
depending on the wave amplitude. 
This wave equation has a variational principle 
and admits local \conslaw/s for energy and momentum 
arising by Noether's theorem
from time- and space- invariance of its corresponding action. 
By applying our \conslaw/ determining system,
we classify all wave speeds for which there are 
extra \conslaw/s of first-order
and obtain the resulting conserved quantities. 
This example illustrates a general approach for classifying 
nonlinear evolution PDEs that admit extra \conslaw/s. 

The third example is a general class of nonlinear Klein-Gordon equations.
The class includes the 
sine-Gordon equation, Liouville equation, and Tzetzeica equation,
which are known to be integrable equations \cite{integrable}
with local \conslaw/s up to arbitrarily high orders,
starting at first order. 
Through our \conslaw/ method 
we give a classification of nonlinear Klein-Gordon equations
admitting at least one second-order \conslaw/.
As a by-product we obtain an integrability characterization 
for some of the equations in this class.
This example shows a general approach to classifying integrable PDEs
by means of \conslaw/s.

\subsection{Generalized Korteweg-de Vries equations}
\label{kdvexample}

Consider the generalized Korteweg-de Vries equation
\EQ\label{kdveq}
\G{}= \Du{t} + u^n\Du{x}+ \Du{xxx} =0 
\endEQ
with parameter $n>0$. 
This is a first order evolution PDE
which has no variational principle directly in terms of $u$
and which reduces for $n=1,2$ to the ordinary KdV equation
and modified KdV equation, respectively. 
Its symmetries 
with infinitesimal generator $\X{}u=\symm{}$ 
\cite{blumanbook,olverbook}
satisfy the determining equation
\EQ\label{kdvsymmeq}
0=\D{t}\symm{} +u^n\D{x}\symm{} +nu^{n-1}\Du{x}\symm{} +\Dx{3}\symm{}
\quad\eqtext{when\quad $G=0$}
\endEQ
where 
$\D{t}=\der{t}+\Du{t}\der{u}+\Du{tx}\der{\Du{x}}+\Du{tt}\der{\Du{t}}+\cdots$ 
and 
$\D{x}=\der{x}+\Du{x}\der{u}+\Du{xx}\der{\Du{x}}+\Du{tx}\der{\Du{t}}+\cdots$ 
are total derivative operators
with respect to $t$ and $x$.
The adjoint of \Eqref{kdvsymmeq} is given by 
\EQ\label{kdvadsymmeq}
0=-\D{t}\adsymm{} -u^n\D{x}\adsymm{}  -\Dx{3}\adsymm{}
\quad\eqtext{when\quad $G=0$}
\endEQ   
which is the determining equation for the adjoint symmetries $\adsymm{}$ of 
the generalized KdV equation. 
(Note, since these determining equations are not self-adjoint,
the only solution common to \Eqrefs{kdvsymmeq}{kdvadsymmeq} is
$\adsymm{}=\symm{}=0$.)

The generalized KdV equation \eqref{kdveq}
itself has the form of a local \conslaw/
\EQ
\D{t}(u) + \D{x}( \frac{1}{n+1} u^{n+1}+ \Du{xx}) =0 . 
\endEQ
We now consider, more generally, local \conslaw/s  
\EQ\label{kdvconslaw}
\D{t}\P{t} + \D{x}\P{x} =0
\endEQ
on all solutions $u(t,x)$ of \Eqref{kdveq}. 
Clearly, we are free without loss of generality
to eliminate any dependence on $\Du{t}$ (and differential consequences)
in the conserved densities $\P{t},\P{x}$.
All nontrivial conserved densities in this form
can be constructed from multipliers $\factor{}$ 
on the generalized KdV equation,
analogous to integrating factors, 
where $\factor{}$ depends only on $t,x,u$, and $x$ derivatives of $u$. 
In particular, 
by moving off the generalized KdV solution space, 
we have
\EQ
\D{t}\P{t} + \D{x}\P{x} = 
( \Du{t} + u^n\Du{x}+ \Du{xxx} ) \Dfactor{0}{}
+ \D{x}( \Du{t} + u^n\Du{x}+ \Du{xxx} ) \Dfactor{1}{}
+ \cdots
\endEQ
for some expressions $\Dfactor{0}{}, \Dfactor{1}{}, \ldots$
with no dependence on $\Du{t}$ and differential consequences. 
This yields (after integration by parts)
the multiplier
\EQ
\D{t}\P{t} + \D{x}(\P{x}-\Gamma) = 
( \Du{t} + u^n\Du{x}+ \Du{xxx} ) \factor{}, \quad
\factor{} = \Dfactor{0}{} - \D{x}\Dfactor{1}{}{}+ \cdots
\endEQ
where $\Gamma=0$ when $u$ is restricted to be a generalized KdV solution. 
We now derive an augmented adjoint symmetry determining system
which completely characterizes all multipliers $\factor{}$. 

The definition for multipliers 
$\factor{}(t,x,u,\deru{}{x},\ldots,\nderu{}{x}{p})$ is that
$(\Du{t} + u^n\Du{x} +\Du{xxx})\factor{}$
must be a divergence expression for all functions $u(t,x)$
(not just generalized KdV solutions). 
This determining condition is expressed by 
\EQs
0= &&
\elop{u}( (\Du{t} + u^n\Du{x} +\Du{xxx})\factor{} ) 
\nonumber\\
= &&
-\D{t}\factor{} -u^n\D{x}\factor{} -\Dx{3}\factor{}
+ (\Du{t} + u^n\Du{x} +\Du{xxx}) \Dfactor{}{u}
-\D{x}( (\Du{t} + u^n\Du{x} +\Du{xxx}) \Dfactor{}{\deru{}{x}} )
\nonumber\\&& \fewquad
+\cdots
+(-1)^p \Dx{p}(  (\Du{t} + u^n\Du{x} +\Du{xxx}) \Dfactor{}{\nderu{}{x}{p}} )
\label{kdvmultiplierdeteq}
\endEQs
where $\elop{u}=
\parderU{}{} -\D{t} \parderU{}{t} -\D{x} \parderU{}{x} 
+\D{t}\D{x}\parderU{}{tx} +\nD{2}{x}\parderU{}{xx}+\cdots$
is the standard Euler operator
which annihilates divergence expressions. 
Equation \eqref{kdvmultiplierdeteq} 
is linear in $\Du{t},\Du{tx},\Du{txx},\ldots,$
and thus the coefficients of $\Du{t}$ and $x$ derivatives of $\Du{t}$ 
up to order $p$
give rise to a split system of
determining equations for $\factor{}$.
The system is found to consist of
the adjoint symmetry determining equation on $\factor{}$,
\EQ\label{kdvadsymmsys}
0=-\solD{t}\factor{} -u^n\D{x}\factor{} -\Dx{3}\factor{}
\endEQ
and extra determining equations on $\factor{}$,
\EQs
&& 0= 
\sum_{k=1}^p (-\Dx{})^{k}\Dfactor{}{\nderu{}{x}{k}} 
\nonumber\\
&& 0= 
(1-(-1)^q) \Dfactor{}{\nderu{}{x}{q}} 
+ \sum_{k=q+1}^{p} \frac{k!}{q!(k-q)!} 
(-\Dx{})^{k-q}\Dfactor{}{\nderu{}{x}{k}} , \quad
q=1,\ldots,p-1
\nonumber\\
&& 0= 
(1-(-1)^p) \Dfactor{}{\nderu{}{x}{p}} . 
\nonumber\\
\label{kdvextrasys}
\endEQs
Here $\solD{t}=
\der{t}-(u\Du{x}+\Du{xxx})\der{u} -(u\Du{x}+\Du{xxx})_x\der{\Du{x}}+\cdots$ 
is the total derivative operator 
which expresses $t$ derivatives of $u$ 
through the generalized KdV equation \eqref{kdveq}. 
Consequently, 
one is able to work on the space of generalized KdV solutions $u(t,x)$ 
in order to solve the determining system 
\eqrefs{kdvadsymmsys}{kdvextrasys} 
to find 
$\factor{}(t,x,u,\deru{}{x},\ldots,\nderu{}{x}{p})$.
The determining system solutions are the multipliers that yield 
all nontrivial generalized KdV \conslaw/s. 

The explicit relation between 
multipliers $\factor{}$ and conserved densities $\P{t},\P{x}$
for generalized KdV \conslaw/s is summarized as follows.
Given a conserved density $\P{t}$, 
we find that a direct calculation of the $\Du{t}$ terms in 
$\D{t}\P{t}+\D{x}\P{x}$ 
yields the multiplier equation
\EQ\label{kdvmultipliereq}
\D{t}\P{t}+\D{x}\P{x} = 
( \Du{t} + u^n\Du{x}+ \Du{xxx} ) \elophat{u}(\P{t})
+ \D{x} \Gamma
\endEQ
where 
$\elophat{u}= 
\parderU{}{} -\D{x} \parderU{}{x} +\nD{2}{x}\parderU{}{xx}+\cdots$
is a restricted Euler operator
and $\Gamma$ is proportional to 
$\Du{t} + u^n\Du{x}+ \Du{xxx}$ and differential consequences. 
Thus we obtain the multiplier
\EQ\label{kdvmultiplier}
\factor{} = \elophat{u}(\P{t}) . 
\endEQ
Conversely, given a multiplier $\factor{}$, 
we can invert the relation \eqref{kdvmultiplier}
by a standard method \cite{olverbook} 
using \Eqref{kdvmultipliereq}
to obtain the conserved density
\EQ\label{kdvPt}
\P{t} = \int_0^1 d\lambda\ u 
\factor{}(t,x,\lambda u,\lambda \deru{}{x},\lambda \nderu{}{x}{2},\ldots). 
\endEQ
From \Eqrefs{kdvmultiplier}{kdvPt}
it is natural to define the order of a generalized KdV \conslaw/ 
as the order of the highest $x$ derivative of $u$ in its multiplier. 
Thus, we see that all nontrivial generalized KdV \conslaw/s up to order $p$
are determined by multipliers of order $p$ 
which are obtained as solutions of 
the augmented system of adjoint symmetry determining equations
\eqrefs{kdvadsymmsys}{kdvextrasys}. 

Through the determining system \eqrefs{kdvadsymmsys}{kdvextrasys} 
we now derive a {\it complete} classification of 
all \conslaw/s \eqref{kdvconslaw} up to second order, 
corresponding to multipliers of the form
\EQ\label{kdvmultform}
\factor{}(t,x,u,\Du{x},\Du{xx}) . 
\endEQ
The classification results are summarized by the following theorem.

\Thm\label{A1}
The generalized KdV equation \eqref{kdveq} for all $n>0$
admits the multipliers
\EQ
\l=1,\ 
\l=u,\ 
\l=\Du{xx} + \frac{1}{n+1} u^{n+1} . 
\label{nkdvmultiplier}
\endEQ
The only additional admitted multipliers of the form \eqref{kdvmultform} 
are given by 
\EQs
&& \l=tu-x, \eqtext{ if $n=1$, } 
\label{nkdvextramultiplier}\\
&& \l=t(\Du{xx} + \frac{1}{3} u^3) - \frac{1}{3} x u, \eqtext{ if $n=2$. } 
\label{nkdvextramultiplier'}
\endEQs
\endThm
This classifies all nontrivial \conslaw/s up to second order
for the generalized KdV equation for any $n>0$. 

The conserved densities for these \conslaw/s 
are easily obtained using the construction formula \eqref{kdvPt} as follows. 
For the multipliers \eqref{nkdvmultiplier}
we find, respectively, 
\EQs
&& \P{t} = u,
\\
&& \P{t} = \frac{1}{2} u^2,
\\
&& \P{t} = -\frac{1}{2} \Du{x}{}^2 + \frac{1}{(n+1)(n+2)} u^{n+2} +\D{x}\theta,
\endEQs
where $\D{x}\theta$ is a trivial conserved density. 
In physical terms, 
if we regard $u$ as a wave amplitude
as in the ordinary ($n=1$) KdV equation,
then these conserved densities represent mass, momentum, and energy
\cite{whitham}. 

For the additional multipliers 
\eqrefs{nkdvextramultiplier}{nkdvextramultiplier'}
we find
\EQs
&& \P{t} = \frac{1}{2} t u^2 - xu,
\eqtext{ if $n=1$, } 
\\
&& \P{t} = -\frac{1}{2} t \Du{x}{}^2 + \frac{1}{12} t u^{4} 
-\frac{1}{6} x u^{2} +\D{x}\theta, 
\eqtext{ if $n=2$. }
\endEQs

\Proof[Proof of Theorem \ref{A1}:]
For multipliers of the form \eqref{kdvmultform}
the adjoint symmetry equation \eqref{kdvadsymmsys} becomes
\EQs
0= &&
-\Dx{3}\l -u^n\D{x}\l -\lder{t} 
+\lder{u} (u^n\Du{x}+\Du{xxx}) 
+\lder{\Du{x}} (u^n\Du{xx}+nu^{n-1}\Du{x}{}^2+\Du{xxxx}) 
\nonumber\\&&\qquad
+\lder{\Du{xx}} (u^n\Du{xxx}+3nu^{n-1}\Du{x}\Du{xx}
+n(n-1)u^{n-2}\Du{x}{}^3 +\Du{xxxxx}) 
\label{Aonead} 
\endEQs
and the extra equations \eqref{kdvextrasys} reduce to 
\EQs
&& 0= -\Dx{}\lder{\Du{x}} +\Dx{2}\lder{\Du{xx}} , 
\label{Aonecon'}\\
&& 0= \lder{\Du{x}} -\D{x}\lder{\Du{xx}} .
\label{Aonecon}
\endEQs
Note that 
\Eqref{Aonecon'} is a differential consequence of \Eqref{Aonecon}.

We start from equation \eqref{Aonecon}.
Its term with highest order derivatives is $\Du{xxx}\lder{\Du{xx}\Du{xx}}$,
and hence $\lder{\Du{xx}\Du{xx}}=0$. 
This yields that $\l$ is linear in $\Du{xx}$,
\EQ\label{Aonel}
\l=a(t,x,u,\Du{x}) \Du{xx} + b(t,x,u,\Du{x}) . 
\endEQ
Then the remaining terms in \Eqref{Aonecon},
after some cancellations, are of first order
\EQ\label{Aoneab}
0=b_{\Du{x}} -a_{u} \Du{x} -a_{x} . 
\endEQ

We now turn to the adjoint symmetry equation \eqref{Aonead}
and separate it into highest derivative terms in descending order. 
The highest order terms cancel.
The second highest order terms involve $\Du{xxxx}$, and these yield
\EQ\label{Aonea}
0=\D{x}a= a_{x} +a_{u}\Du{x} +a_{\Du{x}} \Du{xx} . 
\endEQ
Clearly, 
\Eqref{Aonea} separates into
\EQ\label{Aoneasol}
a_{\Du{x}}=a_{u}=a_{x}=0 . 
\endEQ
Hence $b_{\Du{x}}=0$ from \Eqref{Aoneab}.

The next highest order terms remaining 
in the adjoint symmetry equation \eqref{Aonead}
involve $\Du{xx}$. 
These terms yield
\EQ
0=-a_t-3b_{xu} +3(nu^{n-1} a-b_{uu})\Du{x}
\endEQ
which separates into $b_{uu}= nu^{n-1} a$ and $a_t=-3 b_{xu}$.
Hence we have
\EQ\label{Aoneb}
b=\frac{1}{n+1} au^{n+1}+(b_1(t)-\frac{1}{3}xa_t)u +b_2(t,x) . 
\endEQ
This simplifies the remaining terms in \Eqref{Aonead},
\EQ\label{Aoneab'}
0=-b_{2xxx}-b_{2t} -(b_{1t}-\frac{1}{3} x a_{tt})u 
+b_{2x} u^n +(\frac{1}{3}-\frac{1}{n+1}) a_t u^{n+1} . 
\endEQ
Clearly, the terms here can be separated, yielding
\EQs
&& (n-2) a_t=0 , 
\label{Aoneat}\\
&&  
b_{2t}=-b_{2xxx} , 
\label{Aonebtx}\\
&&  
b_{1t}-\frac{1}{3} x a_{tt} = 
-b_{2x} u^{n-1} .  
\label{Aonebatx}
\endEQs
From \Eqref{Aonebatx}
there are two cases to consider. 

\proclaim{ Case (i):\ $n=1$ }
Here \Eqref{Aoneat} yields 
\EQ\label{Aoneiasol'}
a_t=0
\endEQ
and hence \Eqref{Aonebatx} becomes 
$b_{1t} = -b_{2x}$. 
Since $b_{1}$ depends only on $t$, 
using \Eqref{Aonebtx} we find $b_{2xx}=b_{2t}=0$,
and thus 
\EQ\label{Aoneibsol}
b_1=b_3 t +b_4,\ b_2=-b_3 x+b_5,\ b_3=\const,\ b_4=\const,\ b_5=\const . 
\endEQ
Finally, from \Eqrefs{Aoneasol}{Aoneiasol'} 
it follows that
\EQ\label{Aoneiasolall}
a=\const . 
\endEQ

Therefore, 
through \Eqref{Aonel}, \Eqref{Aoneb}, \Eqrefs{Aoneibsol}{Aoneiasolall},
we have
\EQ
\l = \frac{1}{2} au^2 + a \Du{xx} + b_3 -b_5 x +(b_4+b_5 t) u, 
\eqtext{ if $n=1$, } 
\endEQ
which is a linear combination of the multipliers
\eqrefs{nkdvmultiplier}{nkdvextramultiplier}
shown in Theorem~\ref{A1}.

\proclaim{ Case (ii):\  $n>1$ }
In this case \Eqref{Aonebatx} clearly splits into 
\EQs
&& b_{1t}=a_{tt}=0 , 
\label{Asoliibat}\\
&& b_{2x} =0 . 
\label{Asoliibx}
\endEQs
Hence from \Eqref{Asoliibx} and \Eqref{Aonebtx} 
we have
\EQ\label{Aoneiibsol}
b_2 =\const , 
\endEQ
and from \Eqref{Asoliibat} and \Eqref{Aoneasol}
we have
\EQs
&& a=a_1 t +a_2, a_1 =\const, a_2=\const , 
\label{Aoneiiasol}\\
&& b_1=\const . 
\label{Aoneiibsol'}
\endEQs
Therefore \Eqref{Aoneb} becomes
\EQ\label{Aoneiib}
b=\frac{1}{n+1} (a_1 t + a_2) u^{n+1}
+(b_1-\frac{1}{3}xa_1)u +b_2 . 
\endEQ

Finally, consider \Eqref{Aoneat}. 
If $n\neq 2$ then we have $a_t=0$ and hence $a_1=0$. 
Otherwise, if $n=2$ then we have no restriction on $a_1$. 
Consequently, through 
\Eqref{Aonel}, \Eqrefs{Aoneiiasol}{Aoneiib}, 
we find
\EQ
\l= (a_1 t + a_2)\Du{xx} + \frac{1}{n+1} u^{n+1}(a_1 t + a_2) 
+(b_1-\frac{1}{3} x a_1)u +b_2, 
\eqtext{ if $n>1$, }
\endEQ
where $a_1=0$ for all $n >2$. 
This yields a linear combination of the multipliers 
\eqrefs{nkdvmultiplier}{nkdvextramultiplier'}
shown in Theorem~\ref{A1}.
\endProof

\subsection{Nonlinear wave-speed equation}
\label{wsexample}

Consider the wave equation
\EQ\label{wseq}
\G{}=\Du{tt} -c(u) ( c(u)\Du{x} )_x=0
\endEQ
with a non-constant wave speed $c(u)$.
This is a scalar second-order evolution PDE, 
which has a variational principle given by the physically-motivated action 
\EQ\label{wsaction}
S=\int \frac{1}{2} ( -\Du{t}{}^2 +c(u)^2 \Du{x}{}^2 )dt dx . 
\endEQ
Symmetries of the wave equation \eqref{wseq} 
with infinitesimal generator $\X{}u=\symm{}$
\cite{blumanbook,olverbook}
satisfy the determining equation
\EQs
&&
0=\nD{2}{t}\symm{} -c(u)^2 \nD{2}{x}\symm{} 
-2c(u) c'(u) \Du{x} \D{x}\symm{}
-2c(u) c'(u)\Du{xx}\symm{} 
\nonumber\\&&\qquad
-(c(u)c''(u)+c'(u)^2)\Du{x}{}^2 \symm{} 
\quad\eqtext{when\quad $G=0$.}
\label{wssymmeq}
\endEQs
Since \Eqref{wseq} is variational, 
the determining equation \eqref{wssymmeq} is self-adjoint
and hence the adjoint symmetries of the wave equation 
are the solutions of \Eqref{wssymmeq}. 

For any wave speed $c(u)$,
the wave equation \eqref{wseq} clearly admits 
time and space translation symmetries
\EQ
\symm{} = \Du{t}, \symm{}=\Du{x}, 
\endEQ
which are symmetries of the action. 
In particular, the action is invariant up to a boundary-term
\EQ
\X{}S =\int \frac{1}{2}\D{t}( -\Du{t}{}^2 +c(u)^2 \Du{x}{}^2 )dt dx
\endEQ
under time-translation $\X{}u=\Du{t}$,
and 
\EQ
\X{}S =\int \frac{1}{2} \D{x}( -\Du{t}{}^2 +c(u)^2 \Du{x}{}^2 )dt dx
\endEQ 
under space-translation $\X{}u=\Du{x}$.
By Noether's theorem,
combining the invariance of the action 
with the general variational identity
\EQ
\X{}S = \int\Big( (\Du{tt}-c^2 \Du{xx} -cc'\Du{x}{}^2)\symm{}
+\D{t}(-\Du{t}\symm{}) +\D{x}(c^2\Du{x}\symm{}) \Big) dt dx, \quad
\X{}u=\symm{} , 
\endEQ
we obtain corresponding local \conslaw/s
$\D{t} \P{t} +\D{x} \P{x} =0$
on all solutions $u(t,x)$ of the wave equation \eqref{wseq}.
The conserved densities $\P{t},\P{x}$ are given by 
\EQ\label{wsenergy}
\P{t}= \frac{1}{2} \Du{t}{}^2 +\frac{1}{2} c(u)^2\Du{x}{}^2, 
\P{x}= - c(u)^2 \Du{x}\Du{t}
\endEQ
for the time-translation,
and
\EQ\label{wsmomentum}
\P{t}=  \Du{x}\Du{t},
\P{x}= -\frac{1}{2} \Du{t}{}^2 -\frac{1}{2} c(u)^2\Du{x}{}^2
\endEQ 
for the space-translation.
These yield conservation of energy and momentum, respectively.

We now consider classifying wave speeds $c(u)$ that lead to 
additional local \conslaw/s of first-order
for the wave equation \eqref{wseq}.
(Through Noether's theorem 
all first-order \conslaw/s correspond to 
invariance of the action $S$ under contact symmetries.)

For any local \conslaw/s 
\EQ\label{wsconslaw}
\D{t} \P{t} +\D{x} \P{x} =0
\endEQ
on all solutions $u(t,x)$ of the wave equation \eqref{wseq},
we are clearly free without loss of generality 
to eliminate any dependence on $\Du{tt}$ (and differential consequences)
in the conserved densities $\P{t},\P{x}$.
All nontrivial conserved densities in this form 
can be constructed from multipliers $\factor{}$ on the wave equation,
analogous to integrating factors,
where $\factor{}$ has no dependence on $\Du{tt}$ 
and its differential consequences.
In particular, 
by moving off the wave equation solution space, 
we have 
\EQ
\D{t}\P{t} + \D{x}\P{x} = 
(\Du{tt} -c(u) ( c(u)\Du{x} )_x ) \Dfactor{0}{}
+ \D{x}( \Du{tt} -c(u) ( c(u)\Du{x} )_x ) \Dfactor{1}{}
+ \cdots
\endEQ
for some expressions $\Dfactor{0}{}, \Dfactor{1}{}, \ldots$
with no dependence on $\Du{tt}$ and differential consequences. 
This yields (through integration by parts)
the multiplier
\EQ
\D{t}\P{t} + \D{x}(\P{x}-\Gamma) = 
( \Du{tt} -c(u) ( c(u)\Du{x} )_x ) \factor{}, \quad
\factor{} = \Dfactor{0}{} - \D{x}\Dfactor{1}{}+ \cdots
\endEQ
where $\Gamma=0$ when $u$ is restricted to be a wave equation solution. 
We now derive an augmented symmetry determining system
which completely characterizes all multipliers $\factor{}$.

Multipliers $\factor{}$ are defined by the condition that
$(\Du{tt}-c(u)(c(u)\Du{x})_x) \factor{}$
is a divergence expression for all functions $u(t,x)$
(not just wave equation solutions).
We restrict attention to $\factor{}$ of first-order,
depending on $t,x,u,\Du{t},\Du{x}$.
This leads to the necessary and sufficient determining condition 
\EQs
0 =&& 
\elop{u}( (\Du{tt}-c(u)(c(u)\Du{x})_x) \factor{} )
\nonumber\\
= &&
\nD{2}{t}\factor{} 
-c^2 \nD{2}{x}\factor{}
-2c c' \Du{x} \D{x}\factor{}
-2c c'\Du{xx}\factor{}
-(cc''+c'{}^2)\Du{x}{}^2 \factor{}
\nonumber\\&&\fewquad
+(\Du{tt}-c^2\Du{xx}-cc'\Du{x}{}^2) \Dfactor{}{u}
-\D{t}( (\Du{tt}-c^2\Du{xx}-cc'\Du{x}{}^2) \Dfactor{}{\Du{t}} )
\nonumber\\&&\fewquad
-\D{x}( (\Du{tt}-c^2\Du{xx}-cc'\Du{x}{}^2) \Dfactor{}{\Du{x}} ) 
\label{wsmultiplierdeteq}
\endEQs
where $\elop{u} =
\parderU{}{} -\D{t} \parderU{}{t} -\D{x} \parderU{}{x} 
+\D{t}\D{x}\parderU{}{tx} +\nD{2}{x}\parderU{}{xx}+\cdots$
is the standard Euler operator
which annihilates divergence expressions. 
Equation \eqref{wsmultiplierdeteq}
is quadratic in $\Du{tt}$ and linear in $\Du{ttt}$ and $\Du{ttx}$,
and thus splits into separate equations.
We organize the splitting 
by considering 
terms in $\G{}{}^2$, $\G{}$, $\D{x}\G{}$, $\D{t}\G{}$,
and remaining terms with no dependence on $\Du{tt}$ 
and differential consequences.
The coefficients of $\D{x}\G{}$, $\D{t}\G{}$, and $\G{}{}^2$ 
are found to vanish
as a result of $\factor{}(t,x,u,\Du{t},\Du{x})$ being first-order.
The other coefficients in the splitting do not vanish.

This leads to a split system of two determining equations for 
$\factor{}(t,x,u,\Du{t},\Du{x})$,
consisting of 
\EQ\label{wsadsys}
0=
\solnD{2}{t}\factor{}
-c(u)^2 \nD{2}{x}\factor{}
-2c(u) c'(u) \Du{x} \D{x}\factor{}
-2c(u) c'(u) \Du{xx}\factor{}
-(c(u)c''(u)+c'(u){}^2)\Du{x}{}^2 \factor{}  
\endEQ
which is the symmetry determining equation \eqref{wssymmeq} on $\factor{}$,
and 
\EQ\label{wsextrasys}
0=2\Dfactor{}{u} +\solD{t} \Dfactor{}{\Du{t}} -\D{x}\Dfactor{}{\Du{x}}
\endEQ
which is an extra determining equation on $\factor{}$.
Here $\solD{t}=
\der{t}+\Du{t}\der{u}+\Du{tx}\der{\Du{x}}+ 
(c(u)^2\Du{xx} +c(u)c'(u)\Du{x}{}^2)\der{\Du{t}}+\cdots$
is the total derivative operator 
which expresses $t$ derivatives through the wave equation \eqref{wseq}. 
Consequently, 
one is able to work on the space of wave equation solutions $u(t,x)$ 
in order to solve the determining system 
\eqrefs{wsadsys}{wsextrasys}
to find $\factor{}(t,x,u,\Du{t},\Du{x})$.
The determining system solutions are the multipliers that yield 
all nontrivial first-order \conslaw/s of the wave equation \eqref{wseq}.

The first determining equation \eqref{wsadsys}
shows that all multipliers are symmetries of the wave equation,
while the second determining equation \eqref{wsextrasys}
provides the necessary and sufficient condition for a symmetry 
to leave the action \eqref{wsaction} invariant up to a boundary term.
This is a consequence of 
the one-to-one correspondence between symmetries of the action
and multipliers for nontrivial \conslaw/s of 
the wave equation \eqref{wseq} 
as shown by Noether's theorem \cite{olverbook}. 

The explicit relation between 
multipliers $\factor{}$ and conserved densities $\P{t},\P{x}$
for first-order \conslaw/s \eqref{wsconslaw}
is summarized as follows.
Given a conserved density $\P{t}$, 
we find that a direct calculation of the $\Du{tt}$ terms in 
$\D{t}\P{t}+\D{x}\P{x}$ 
yields the multiplier equation
\EQ\label{wsmultipliereq}
\D{t}\P{t}+\D{x}\P{x} = 
(\Du{tt} -c(u) ( c(u)\Du{x} )_x ) \elophat{\Du{t}}(\P{t})
+ \D{x} \Gamma
\endEQ
where $\elophat{\Du{t}}= \parderU{}{t}$
is the truncation of a restricted Euler operator, 
and $\Gamma$ is proportional to 
$\Du{tt} -c(u) ( c(u)\Du{x} )_x$ and differential consequences. 
Thus we obtain the multiplier
\EQ\label{wsmultiplier}
\factor{}=\elophat{\Du{t}} (\P{t}) . 
\endEQ
Conversely, given a multiplier $\factor{}$, 
we can invert the relation \eqref{wsmultiplier}
using \Eqref{wsmultipliereq}
to obtain the conserved density
\EQ\label{wsPt}
\P{t} = 
\int_0^1 d\lambda \Big(
(\Du{t} -\Dtu{t})\factor{}[\lambda \u{}+(1-\lambda)\tu{}]
+(\tu{}-\u{})\solD{t}\factor{}[\lambda \u{}+(1-\lambda)\tu{}] \Big)
+t\int_0^1 d\lambda  \K(\lambda t,\lambda x)
\endEQ
where 
\EQs
&& \factor{}[u]=\factor{}(t,x,u,\Du{t},\Du{x}), 
\\
&& \solD{t}\factor{}[u]= 
\Dfactor{}{t}(t,x,u,\Du{t},\Du{x})
+ \Du{t}\Dfactor{}{u}(t,x,u,\Du{t},\Du{x})
+ \Du{tx}\Dfactor{}{\Du{x}}(t,x,u,\Du{t},\Du{x})
\nonumber\\&&\manyquad
+ c(u)(c(u)\Du{x})_x \Dfactor{}{\Du{t}}(t,x,u,\Du{t},\Du{x}) , 
\\
&& \K(t,x) = 
(\Dtu{tt} -c(\tu{})( c(\tu{})\Dtu{x} )_x )\factor{}[\tu{}], 
\endEQs
with $\tu{}=\tu{}(t,x)$ being any function chosen such that
the expressions $\factor{}[\tu{}]$ 
and $\Dtu{tt} -c(\tu{})( c(\tu{})\Dtu{x} )_x$
are non-singular. 

Thus we see that all nontrivial first-order \conslaw/s of
the wave equation \eqref{wseq}
are determined by multipliers of first-order
which are obtained as solutions of 
the augmented system of symmetry determining equations
\eqrefs{wsadsys}{wsextrasys}.

We now use the determining system \eqrefs{wsadsys}{wsextrasys} 
to {\it completely} classify all first-order \conslaw/s
in terms of corresponding multipliers 
\EQ\label{wsmultiplierclass}
\factor{}(t,x,u,\Du{t},\Du{x}) . 
\endEQ
The classification results are summarized by the following two theorems.

\Thm\label{B1}
For arbitrary wave speeds $c(u)$, 
the only multipliers of form \eqref{wsmultiplierclass}
admitted by the wave equation \eqref{wseq} are
\EQ\label{wsmultipliers}
\factor{}=\Du{t} , 
\factor{}=\Du{x} ,
\factor{}=t\Du{t}+x\Du{x} . 
\endEQ 
\endThm

These multipliers define symmetries given by 
time-translation, space-translation, and time-space dilation, respectively,
which lead to \conslaw/s for 
energy \eqref{wsenergy}, 
momentum \eqref{wsmomentum},
and a dilational quantity. 

\Thm\label{B2}
The wave equation \eqref{wseq} for non-constant wave speed $c(u)$
admits additional multipliers of form \eqref{wsmultiplierclass}
iff $c(u)=c_0 (u-u_0)^{-2}$ in terms of constants $c_0,u_0$. 
For these wave speeds the additional admitted multipliers are given by 
\EQs
&& \factor{} =t^2\Du{t} -t(u-u_0) , 
\label{wstconformal}\\
&& \factor{} = x^2\Du{x} +x(u-u_0) , 
\label{wsxconformal}\\
&& \factor{} = t\Du{t} -x\Du{x} -(u-u_0) . 
\label{wsboost}
\endEQs 
\endThm

The symmetries defined by these additional multipliers correspond to 
two conformal (M\"obius) transformations and one scaling transformation
on independent variables $t,x$, 
accompanied by a scaling and shift of $u$.
The conserved densities $\P{t}$ for the three corresponding \conslaw/s
are obtained by the construction formula \eqref{wsPt}.
We choose $\tu{}=\const \neq u_0$ to avoid the singularity at $u=u_0$. 
This leads to $\K=0$.
Using the linear dependence of $\factor{}(t,x,u,\Du{x},\Du{t})$
together with the property $\Dfactor{}{x\Du{t}}=0$
given by \Eqsref{wstconformal}{wsboost}, 
we find that after integration by parts 
the formula reduces to 
\EQs
\P{t} &&
= \frac{1}{2} \Du{t} \Big(
\factor{}[u] + \factor{}[\tu{}] 
+ ( (\u{}-\tu{})\Dfactor{}{\Du{x}}[\tu{}] )_x \Big)
+ \frac{1}{2} (\tu{}-\u{}) \Big(
\Dfactor{}{t}[u] + \Dfactor{}{t}[\tu{}] 
+ \Du{t} \Dfactor{}{u}[\tu{}] \Big)
\nonumber\\&&\fewquad
+\frac{1}{2} \Du{x}{}^2 c(\u{})^2 \Dfactor{}{\Du{t}}[\tu{}]
\label{wsPt'}
\endEQs
up to a trivial conserved density $\D{x}\theta$.
Since the terms in \Eqref{wsPt'} are non-singular for any constant $\tu{}$,
we can now set $\tu{}=u_0$ without loss of generality.
Then, substituting \Eqsref{wstconformal}{wsboost} for $\factor{}$, 
we obtain respectively
\EQs
&&
\P{t} = \frac{1}{2} t^2 \Du{t}{}^2 -t(u-u_0) \Du{t}
+\frac{1}{2} (u-u_0)^2 +\frac{1}{2} t^2 \Du{x}{}^2 c_0{}^2 (u-u_0)^{-4} , 
\\
&&
\P{t} = x^2 \Du{x}\Du{t} + x(u-u_0) \Du{t} , 
\\
&&
\P{t} = \frac{1}{2} t \Du{t}{}^2 - (x\Du{x} +u-u_0) \Du{t}
+\frac{1}{2} t \Du{x}{}^2 c_0{}^2 (u-u_0)^{-4} , 
\endEQs
which represent two conformal quantities and a scaling quantity. 

\Proof[Proof of Theorems \ref{B1} and \ref{B2}:]
We start from the determining equation \eqref{wsextrasys}
and expand it in explicit form using
\EQs
&& \solD{t} \factor{} =
\Dfactor{}{t} +\Dfactor{}{u} \Du{t} + \Dfactor{}{\Du{x}} \Du{tx}
+ \Dfactor{}{\Du{t}} (c^2 \Du{xx} +cc'\Du{x}{}^2),
\\
&& \D{x} \factor{} =
\Dfactor{}{x} +\Dfactor{}{u} \Du{x} + \Dfactor{}{\Du{x}} \Du{xx}
+ \Dfactor{}{\Du{t}} \Du{tx} . 
\endEQs
This yields
\EQ
0 = 
2\Dfactor{}{u} +\Dfactor{}{t\Du{t}}  +\Dfactor{}{u\Du{t}} \Du{t}
+cc'\Du{x}{}^2 \Dfactor{}{\Du{t}\Du{t}} 
-\Dfactor{}{x\Du{x}} -\Dfactor{}{u\Du{x}} \Du{x} 
+(c^2 \Dfactor{}{\Du{t}\Du{t}} -\Dfactor{}{\Du{x}\Du{x}}) \Du{xx} .
\label{Bextra}
\endEQ
Since $\factor{}$ does not depend on $\Du{xx}$,
\Eqref{Bextra} separates into 
\EQs
&& 0=
c^2 \Dfactor{}{\Du{t}\Du{t}} - \Dfactor{}{\Du{x}\Du{x}} , 
\label{Bwave}\\
&& 0= 
2\Dfactor{}{u} +\Dfactor{}{t\Du{t}}  +\Dfactor{}{u\Du{t}} \Du{t} 
-\Dfactor{}{x\Du{x}} -\Dfactor{}{u\Du{x}} \Du{x} . 
\label{Bextra'}
\endEQs
Next we bring in the symmetry determining equation \eqref{wsadsys}.
Expanding it similarly, we find after use of \Eqref{Bwave}
that its highest derivative terms involve 
$\Du{xx}$ and $\Du{tx}$. 
Hence, since $\factor{}$ does not depend on second-derivatives of $u$,
these terms can be separated, leading to
\EQs
&& 0= 
( \Dfactor{}{t\Du{t}}  +\Dfactor{}{u\Du{t}} \Du{t}
-\Dfactor{}{x\Du{x}} -\Dfactor{}{u\Du{x}} \Du{x} ) c
+( \Dfactor{}{\Du{t}} \Du{t} +\Dfactor{}{\Du{x}} \Du{x} -\factor{} ) c'
+\Dfactor{}{\Du{t}\Du{t}} \Du{x}{}^2 c^2c' , 
\nonumber\\&&
\label{Bsymmuxx}\\
&& 0 = 
\Dfactor{}{t\Du{x}}  +\Dfactor{}{u\Du{x}} \Du{t}
-( \Dfactor{}{x\Du{t}} +\Dfactor{}{u\Du{t}} \Du{x} ) c^2
+\Dfactor{}{\Du{x}\Du{t}} \Du{x}{}^2 cc' . 
\label{Bsymmutx}
\endEQs

To proceed, 
we combine \Eqrefs{Bextra'}{Bsymmuxx} to yield
\EQ
0=( \Dfactor{}{\Du{t}} \Du{t} +\Dfactor{}{\Du{x}} \Du{x} -\factor{} ) c' 
-2\Dfactor{}{u} c 
\label{Bfactoreq}
\endEQ
which is a first order linear PDE for $\factor{}$ in $u,\Du{t},\Du{x}$. 
The solution of \Eqref{Bfactoreq} is given by 
\EQ
\factor{}=c^{-1/2} f(t,x,\alpha,\beta), \quad
\alpha = c^{1/2} \Du{t}, \beta = c^{1/2} \Du{x} . 
\label{Bfactorsol}
\endEQ
Substituting \Eqref{Bfactorsol} into \Eqref{Bwave}, we obtain
$c^2 f_{\alpha\alpha} = f_{\beta\beta}$
which separates into 
\EQ
f_{\alpha\alpha} = f_{\beta\beta} =0
\endEQ
since $c(u)\neq \const$. 
Hence we have 
\EQ
f=a(t,x)\alpha +b(t,x)\beta +g(t,x)\alpha\beta +h(t,x) .
\label{Bf}
\endEQ
Next, we substitute \Eqrefs{Bf}{Bfactorsol} into \Eqref{Bsymmutx}
to obtain
\EQ
0= b_t-c^2 a_x +\frac{1}{2} c^{-1/2} c' (\Du{t}{}^2+c^2 \Du{x}{}^2) g .
\endEQ
This immediately yields
\EQ
g=0, b_t=a_x=0 . 
\label{Bgbtax}
\endEQ
Thus, so far we have
\EQ
\factor{} = a(t) \Du{t} + b(x) \Du{x} + h(t,x) c(u)^{-1/2} . 
\label{Bfactor}
\endEQ

Now, using \Eqref{Bfactor} we see that 
\Eqrefs{Bextra'}{Bsymmuxx} both reduce to 
\EQ
0=-c^{-3/2} c' h + a_t -b_x . 
\endEQ
This leads to two cases to consider.
If the wave speed $c(u)$ is arbitrary, then we must have
\proclaim{ Case (i):
\EQ\label{Bcasei}
a_t=b_x, h=0 . 
\endEQ}
Alternatively, 
the only other possibility is for the wave speed $c(u)$ 
to satisfy a first-order ODE, so we then have
\proclaim{ Case (ii):
\EQ\label{Bcaseii}
c^{-3/2} c' =\tc =\const,
\tc h = a_t-b_x . 
\endEQ}
Note in this case we require $\tc\neq 0$
in order for $c(u)$ to be non-constant.

Finally, we return to the symmetry determining equation \eqref{wsadsys}
and consider the terms that remain after we substitute \Eqref{Bfactor}. 
The analysis proceeds according to the two cases.

In case (i), the remaining terms in \Eqref{wsadsys} reduce to 
\EQ\label{Bioth}
0=-c^2 b_{xx} \Du{x} + a_{tt} \Du{t}
\endEQ
which separates into 
\EQ
b_{xx}=a_{tt}=0 . 
\endEQ
Solving this equation and using \Eqrefs{Bcasei}{Bgbtax},
we obtain
\EQ\label{Biab}
a=a_0 +a_1 t,
b=b_0+a_1 x
\endEQ
where $a_0,b_0,a_1$ are constants.
Consequently, from \Eqref{Bfactor}, we have
\EQ\label{Bifactor}
\factor{} = a_0 \Du{t} + b_0 \Du{x} + a_1 (t\Du{t} +x \Du{x}) . 
\endEQ
Hence, these are the only multipliers of first-order 
admitted by the wave equation \eqref{wseq}
for arbitrary wave speeds $c(u)$. 
This establishes Theorem~\ref{B1}.

Finally, in case (ii), 
using \Eqref{Bcaseii} for $h$ and \Eqref{Bgbtax} for $a$ and $b$, 
then eliminating $c'$ by \Eqref{Bcaseii},
we find that the remaining terms in \Eqref{wsadsys} 
simplify considerably to 
\EQ
0= c^{-1/2} h_{tt} -c^{-3/2} h_{xx} . 
\endEQ
Hence, since $c(u)$ is non-constant, we obtain
\EQ\label{Biih}
h_{tt}=h_{xx} =0 . 
\endEQ
Now we solve for $a,b$ by combining \Eqref{Biih} 
with \Eqrefs{Bcaseii}{Bgbtax},
which gives
\EQ
a_{ttt}=b_{xxx}=0 . 
\endEQ
Hence we have
\EQ\label{Biiab}
a= a_0+a_1 t +a_2 t^2,
b=b_0 +b_1 x + b_2 x^2 , 
\endEQ
where $a_0,a_1,a_2,b_0,b_1,b_2$ are constants.
Then \Eqref{Bcaseii} yields
\EQ\label{Biihsol}
\tc h = a_1 +2a_2 t -b_1 -2b_2 x . 
\endEQ
Consequently, from \Eqref{Bfactor}, we obtain
\EQs
\factor{} = &&
a_0 \Du{t} + b_0 \Du{x} 
+ a_1 (t\Du{t} + c^{-1/2}/\tc) +b_1 (x\Du{x} - c^{-1/2}/\tc)
\nonumber\\&&
+a_2 ( t^2 \Du{t} +2t c^{-1/2}/\tc ) +b_2 ( x^2 \Du{x} -2x c^{-1/2}/\tc )
\label{Biifactor}    
\endEQs
where the wave speed is given from \Eqref{Bcaseii} by 
\EQ
c(u) = c_0 (u-u_0)^{-2}
\endEQ
with $\tc=-2 c_0{}^{-1/2}$, $c_0>0$. 

In \Eqref{Biifactor} 
we have a six parameter family of admitted multipliers.
The parameters $a_0,b_0,a_1=b_1$ yield the three multipliers of case (i).
Hence the present case has three additional multipliers
arising from the parameters $a_1=-b_1,a_2,b_2$. 
This establishes Theorem~\ref{B2}.
\endProof

\subsection{Klein-Gordon wave equations}
\label{kleingordonexample}

Consider the class of Klein-Gordon wave equations 
\EQ\label{kgeq}
\G{}=\Du{tx} -g(u) =0
\endEQ
with a nonlinear interaction $g(u)$.
This class has a variational principle given by the action
\EQ\label{kgaction}
S=-\int (\frac{1}{2} \Du{t}\Du{x} + h(u)) dtdx, \quad
h'(u)=g(u) . 
\endEQ
Since the general Klein-Gordon equation \eqref{kgeq} is variational, 
its symmetries with infinitesimal generator $\X{}u=\symm{}$ 
\cite{blumanbook,olverbook}
satisfy the determining equation
\EQ\label{kgsymmeq}
0=\D{t}\D{x}\symm{} -g'(u)\symm{}
\quad\eqtext{when\quad $G=0$}
\endEQ
which is self-adjoint. 
Hence the adjoint symmetries of \Eqref{kgeq} 
are the solutions of \Eqref{kgsymmeq}.

The symmetries $\X{}u=\symm{}$ 
admitted by the general Klein-Gordon equation \eqref{kgeq}
clearly consist of $t$ and $x$ translations
\EQ\label{kgtranslationsymm}
\symm{}=\Du{t},
\symm{}=\Du{x}, 
\endEQ
as well as a $t,x$ boost
\EQ\label{kgboostsymm}
\symm{}=t\Du{t}-x\Du{x} . 
\endEQ
These are easily checked to be symmetries of the action,
leaving $S$ invariant up to a boundary-term, with
\EQ
\X{}S = -\int \D{t}( \frac{1}{2} \Du{t}\Du{x} + h(u) )dtdx,
\quad
\X{}S = -\int \D{x}( \frac{1}{2}  \Du{t}\Du{x} + h(u) )dtdx
\endEQ
under the respective translations
$\X{}u=\Du{t}$ and $\X{}u=\Du{x}$,
and with 
\EQ
\X{}S = \int \Big(
-\D{t}( \frac{1}{2} t\Du{t}\Du{x} + t h(u) )
+\D{x}( \frac{1}{2}  x\Du{t}\Du{x} + x h(u) )
\Big)dtdx
\endEQ
under the boost $\X{}u= t\Du{t}-x\Du{x}$.
By Noether's theorem, 
combining the invariance of the action
and the general variational identity
\EQ
\X{}S = \int \Big(
(\Du{tx} - g(u) )\symm{} - \D{t} (\frac{1}{2} \symm{} \Du{x})
- \D{x} (\frac{1}{2} \symm{} \Du{t})
\Big)dtdx, 
\quad \X{}u=\symm{}, 
\endEQ
we obtain corresponding local \conslaw/s
$\D{t}\P{t} +\D{x}\P{x} =0$
on all solutions $u(t,x)$ of the Klein-Gordon equation \eqref{kgeq}.
The conserved densities are given by 
\EQs
&&
\P{t} = h(u),
\P{x} =-\frac{1}{2} \Du{t}{}^2 , 
\\&&
\P{t} =- \frac{1}{2} \Du{x}{}^2,
\P{x} = h(u) , 
\endEQs
from the translations,
and
\EQ
\P{t} =  \frac{1}{2} x\Du{x}{}^2 +t h(u),
\P{x} =-\frac{1}{2} t\Du{t}{}^2 -x h(u) , 
\endEQ
from the boost.
These comprise all of the first-order local \conslaw/s 
for the general Klein-Gordon equation \eqref{kgeq}.

The Klein-Gordon equations in the class \eqref{kgeq} include
\EQs
&& \Du{tx} = \sin u, \fewquad\eqtext{ sine-Gordon equation }
\label{sgeq}\\
&& \Du{tx} = e^u \pm e^{-u}, \fewquad\eqtext{ sinh-Gordon equation }
\label{shgeq}\\                
&& \Du{tx} = e^u \pm e^{-2u}, \fewquad\eqtext{ Tzetzeica equation }
\label{tzetzeicaeq}\\                
&& \Du{tx} = e^u. \fewquad\eqtext{ Liouville equation }
\label{liouvilleeq}
\endEQs
The first three are soliton equations 
while the last is a linearizable equation.
These are singled out \cite{integrable}
as nonlinear wave equations 
that are known to be integrable in the sense of 
admitting an infinite number of higher-order local \conslaw/s
\cite{kgconslaws,tzetzeicaeq}.
For each equation the \conslaw/s fall into two sequences
where $\P{t}$ and $\P{x}$ depend 
purely on $u$ and $t$ derivatives of $u$ in one sequence,
and purely on $u$ and $x$ derivatives of $u$ in the other sequence
(corresponding to the reflection symmetry $t\leftrightarrow x$).

Here, for the class of nonlinear Klein-Gordon equations \eqref{kgeq}, 
we consider local \conslaw/s of higher-order
\EQ\label{kgconslaw}
\D{t} \P{t} +\D{x} \P{x} =0
\endEQ
with $\P{t},\P{x}$ depending 
either purely on $t,u$, and $t$ derivatives of $u$,
or purely on $x,u$, and $x$ derivatives of $u$, 
on all solutions $u(t,x)$ of \Eqref{kgeq}. 

All nontrivial \conslaw/s \eqref{kgconslaw} 
with conserved densities of the particular form
\EQ\label{kgpuretcons}
\P{t}(x,u,\deru{}{x},\ldots,\nderu{}{x}{q}),
\P{x}(x,u,\deru{}{x},\ldots,\nderu{}{x}{q})
\endEQ
can be constructed from multipliers $\factor{}$ 
on the Klein-Gordon equation \eqref{kgeq}, 
analogous to integrating factors,
with the dependence
\EQ\label{kgpuretmultiplier}
\factor{}(x,u,\deru{}{x},\ldots,\nderu{}{x}{p})
\endEQ
where $p=2q-1$.
In particular, 
moving off the Klein-Gordon solution space, 
we have 
\EQ
\D{t}\P{t} + \D{x}\P{x} = 
(\Du{tx} -g(u) ) \Dfactor{0}{}
+ \D{x}( \Du{tx} -g(u) ) \Dfactor{1}{}
+ \cdots
\endEQ
for some expressions $\Dfactor{0}{}, \Dfactor{1}{}, \ldots$
with no dependence on $\Du{t}$ and differential consequences. 
This yields (after integration by parts) the multiplier
\EQ
\D{t}\P{t} + \D{x}(\P{x}-\Gamma) = 
( \Du{tx} -g(u) ) \factor{}, \quad
\factor{} = \Dfactor{0}{} - \D{x}\Dfactor{1}{}+ \cdots
\endEQ
where $\Gamma=0$ when $u$ is restricted to be a Klein-Gordon solution. 
There are corresponding \conslaw/s of mirror form 
produced by the transformation
$x\leftrightarrow t$, $\nderu{}{x}{k}\leftrightarrow \nderu{}{t}{k}$, 
with $\P{t} \leftrightarrow \P{x}$.
We now derive an augmented symmetry determining system
which completely characterizes the \conslaw/ multipliers 
\eqref{kgpuretmultiplier}.

The determining condition for multipliers $\factor{}$
is that $(\Du{tx} -g(u))\factor{}$ 
must be a divergence expression
for all functions $u(t,x)$ (not just Klein-Gordon solutions).
For multipliers of the form \eqref{kgpuretmultiplier},
this condition is expressed by 
\EQs
0=&& 
\elop{u}( (\Du{tx}-g)\factor{} )
\nonumber\\
= &&
\D{t}\D{x}\factor{} -g'\factor{} 
+\Dfactor{}{u} (\Du{tx}-g)
-\D{x}( \Dfactor{}{\deru{}{x}} (\Du{tx}-g) )
+\cdots
+(-\D{x})^p( \Dfactor{}{\nderu{}{x}{p}} (\Du{tx}-g) ) 
\nonumber\\&&
\label{kgmultiplierdeteq}
\endEQs
where $\elop{u} =
\parderU{}{} -\D{t} \parderU{}{t} -\D{x} \parderU{}{x} 
+\D{t}\D{x}\parderU{}{tx} +\nD{2}{x}\parderU{}{xx}+\nD{2}{t}\parderU{}{tt}
+\cdots$
is the standard Euler operator
which annihilates divergence expressions. 
Equation \eqref{kgmultiplierdeteq}
is linear in $\Du{tx},\Du{txx},\ldots$,
and hence splits into separate equations. 
We organize the splitting 
in terms of $\G{},\D{x}\G{},\ldots,\nD{p}{x}\G{}$.
This yields a split system of determining equations
for $\factor{}(x,u,\deru{}{x},\ldots,\nderu{}{x}{p})$,
which is found to consist of 
the Klein-Gordon symmetry determining equation on $\factor{}$
\EQ\label{kgsymmsys}
0=\solD{t} \D{x}\factor{} -g'(u)\factor{}
\endEQ
and extra determining equations on $\factor{}$
\EQs
&& 0= 2\Dfactor{}{u} +\sum_{k=2}^p (-\D{x})^k \Dfactor{}{\nderu{}{x}{k}} , 
\label{kgextrasysone}\\
&& 0= (1+(-1)^j) \Dfactor{}{\nderu{}{x}{j}} 
+((-1)^j-j-1) \D{x} \Dfactor{}{\nderu{}{x}{j+1}} 
+\sum_{k=j+2}^p  \frac{k!}{j!(k-j)!} 
(-\D{x})^{k-j} \Dfactor{}{\nderu{}{x}{k}} ,
\nonumber\\&&\fewquad
j=1,\ldots,p-1
\label{kgextrasysmore}\\
&& 0=(1+(-1)^p)\Dfactor{}{\nderu{}{x}{p}} . 
\label{kgextrasyslast}
\endEQs
Here $\solD{t} = 
\der{t} +\Du{t}\der{u} +g(u)\der{\Du{x}} +g'(u)\Du{x}\der{\Du{xx}} +\cdots$ 
is the total derivative operator 
which expresses $t$ derivatives of $u$ 
through the Klein-Gordon equation \eqref{kgeq}. 
Consequently, 
one is able to work on the space of Klein-Gordon solutions $u(t,x)$ 
in order to solve the determining system 
\eqsref{kgsymmsys}{kgextrasyslast}
to find $\factor{}(x,u,\deru{}{x},\ldots,\nderu{}{x}{p})$.

The solutions of 
the determining system \eqsref{kgsymmsys}{kgextrasyslast}
are the multipliers \eqref{kgpuretmultiplier}
that yield all nontrivial \conslaw/s 
of the form \eqref{kgpuretcons}.
Noether's theorem establishes that
there is a one-to-one correspondence between symmetries of the action
and multipliers for nontrivial \conslaw/s of the Klein-Gordon equation \eqref{kgeq}.
Consequently, it follows that 
the extra determining equations \eqsref{kgextrasysone}{kgextrasyslast}
represent the necessary and sufficient condition
for a symmetry to leave the action \eqref{kgaction} invariant
up to a boundary term.  

The explicit relation between 
multipliers $\factor{}$ and conserved densities $\P{t},\P{x}$
for \conslaw/s \eqref{kgpuretcons}
is summarized as follows.
Given a conserved density $\P{t}$, 
we find that a direct calculation of the $\Du{tx}$ terms in 
$\D{t}\P{t}+\D{x}\P{x}$ 
yields the multiplier equation
\EQ\label{kgmultipliereq}
\D{t}\P{t}+\D{x}\P{x} = 
( \Du{tx} -g(u) ) \elophat{\Du{x}}(\P{t})
+ \D{x} \Gamma
\endEQ
where 
$\elophat{\Du{x}} = \parderU{}{x} -\D{x}\parderU{}{xx} + \cdots$
is a restricted Euler operator, 
and $\Gamma$ is proportional to 
$\Du{tx} -g(u)$ and differential consequences. 
Thus we obtain the multiplier
\EQ\label{kgmultiplier}
\factor{}=\elophat{\Du{x}}( \P{t} ) . 
\endEQ
Conversely, given a multiplier $\factor{}$, 
we can invert the relation \eqref{kgmultiplier}
using \Eqref{kgmultipliereq}
to obtain the conserved density
\EQ\label{kgPt}
\P{t} = \int_0^1 d\lambda (\Du{x} -\Dtu{x})
\factor{}[\lambda \u{}+(1-\lambda)\tu{}]
\endEQ
where $\factor{}[u]=\factor{}(x,u,\deru{}{x},\ldots,\nderu{}{x}{p})$, 
and $\tu{}=\tu{}(x)$ is any function chosen so that
the expressions $\factor{}[\tu{}]$ and $g(\tu{})$ are non-singular. 
In particular, if $\factor{}[0]$ and $g(0)$ are non-singular,
then we can choose $\tu{}=0$, which simplifies the integral \eqref{kgPt}.

Thus, from \Eqrefs{kgmultiplier}{kgPt}, 
we see that all nontrivial Klein-Gordon \conslaw/s \eqref{kgpuretcons}
of order $q$ are determined by multipliers \eqref{kgpuretmultiplier}
of order $p=2q-1$ 
which are obtained as solutions of 
the augmented system of symmetry determining equations
\eqsref{kgsymmsys}{kgextrasyslast}.

Through the determining system
\eqsref{kgsymmsys}{kgextrasyslast},
we now derive a classification of 
nonlinear Klein-Gordon interactions $g(u)$
that admit higher-order \conslaw/s \eqref{kgpuretcons}
starting at order $q=2$. 
The classification results are summarized by three theorems. 

\Thm\label{C1}
The nonlinear Klein-Gordon equation \eqref{kgeq}
admits a \conslaw/ \eqref{kgpuretcons} of order $q=2$
iff the nonlinear interaction is given by one of 
\EQ
g(u)=e^u, g(u) =\sin u, g(u) = e^u \pm e^{-u}
\label{kginteractions}
\endEQ
to within scaling of $t,x,u$, and translation of $u$. 
\endThm
(If we allow $u$ to undergo complex-valued scalings and translations
then the Klein-Gordon equations for $g(u)=\sin u$ and $g(u)=e^u \pm e^{-u}$ 
are equivalent.)

The Klein-Gordon equations arising from Theorem~\ref{C1} are 
the Liouville equation \eqref{liouvilleeq},
sine-Gordon equation \eqref{sgeq}
and related sinh-Gordon equation \eqref{shgeq}.
(The Tzetzeica equation \eqref{tzetzeicaeq} is absent 
in this classification 
because its first admitted higher-order \conslaw/ \eqref{kgpuretcons} 
is of order $q=3$.)
Since each of these Klein-Gordon equations is known to admit
an infinite sequence of higher order \conslaw/s \eqref{kgpuretcons}
for $q\ge 2$,
this leads to an integrability classification. 

\Cor\label{C2}
A nonlinear Klein-Gordon equation \eqref{kgeq} is integrable
in the sense of having an infinite number of 
\conslaw/s \eqref{kgpuretcons} up to arbitrarily high orders $q>1$
if it admits one of order $q=2$. 
\endCor

\Thm\label{C3}
The multipliers for the second-order \conslaw/s \eqref{kgpuretcons} 
admitted by the Klein-Gordon equation \eqref{kgeq}
with nonlinear interactions \eqref{kginteractions}
are given by 
\EQs
&& \factor{}= \Du{xxx} +\frac{1}{2}\Du{x}{}^3 , 
\label{sgmultiplier}\\
&& \factor{}= \Du{xxx} -\frac{1}{2}\Du{x}{}^3 , 
\label{shgmultiplier}
\endEQs
for the sine-Gordon equation \eqref{sgeq} 
and sinh-Gordon equation \eqref{shgeq}, respectively, 
and 
\EQ
\factor{}= (\Du{xxx}-\Du{x}\Du{xx}) f_\xi +f_x + \Du{x} f
= \D{x}f + \Du{x} f , 
\label{liouvillemultiplier}
\endEQ
depending on an arbitrary function
\EQ
f=f(x,\xi), \quad
\xi= \Du{xx}-\frac{1}{2} \Du{x}{}^2 , 
\label{liouvillemultiplier'}
\endEQ
for the Liouville equation \eqref{liouvilleeq}. 
\endThm

From the construction formula \eqref{kgPt}
the corresponding conserved densities 
for the multipliers \eqrefs{sgmultiplier}{shgmultiplier}, respectively, 
are given by
\EQ
\P{t} = \frac{1}{2} \Du{xxx}\Du{x} \pm \frac{1}{8} \Du{x}{}^4 
=-\frac{1}{2} \Du{xx}{}^2 \pm \frac{1}{8} \Du{x}{}^4 +\D{x}\theta , 
\endEQ
where we have used $\tu{}=0$, since $\factor{}[0]=0$ and $g(0)=\const$.
Similarly, the conserved density 
for the multiplier \eqref{liouvillemultiplier} 
is given by
\EQs
\P{t} &&
= (\Du{x} -\Dtu{x}) \D{x} \int_0^1 d\lambda 
f( x, \lambda(\Du{xx} -\Dtu{xx})-\frac{1}{2}\lambda^2 (\Du{x} -\Dtu{x})^2 )
\nonumber\\&&\fewquad
+ (\Du{x} -\Dtu{x})^2 \int_0^1 d\lambda\ \lambda
f( x, \lambda(\Du{xx} -\Dtu{xx})-\frac{1}{2}\lambda^2 (\Du{x} -\Dtu{x})^2 )
\nonumber\\&&
=  -\int_{\Dtu{xx}-\frac{1}{2}\Dtu{x}{}^2}^{\Du{xx}-\frac{1}{2}\Du{x}{}^2}
f(x,\xi) d\xi +\D{x}\theta , 
\label{liouvillePt}
\endEQs
where now $\tu{}=\tu{}(x)$ is any function chosen so that 
$\factor{}[\tu{}]=\D{x}f(x,\tilde\xi)+\Dtu{x}f(x,\tilde\xi)$ 
and $g(\tu{})=\exp(\tu{})$ are non-singular expressions. 

We remark that the existence of the \conslaw/ \eqref{liouvillePt}
involving an arbitrary function 
reflects the well-known classical integrability \cite{liouvilleeq}
(in the sense of explicit integration) 
of the Liouville equation.

\Proof[Proof of Theorem \ref{C1}:]

From \Eqrefs{kgpuretcons}{kgpuretmultiplier}, 
\conslaw/s with 
$\P{t}(x,u,\Du{x},\Du{xx})$ 
and \break
$\P{x}(x,u,\Du{x},\Du{xx})$
of order $q=2$
correspond to multipliers 
\EQ\label{Cdependence}
\factor{}(x,u,\Du{x},\Du{xx},\Du{xxx}) . 
\endEQ
We start with the extra determining equations 
\eqsref{kgextrasysone}{kgextrasyslast}.
These become
\EQs
&& 0=2\Dfactor{}{u} +\nD{2}{x}\Dfactor{}{\Du{xx}} 
-\nD{3}{x}\Dfactor{}{\Du{xxx}} , 
\label{Cextra}\\
&& 0= \D{x}\Dfactor{}{\Du{xx}} -\nD{2}{x}\Dfactor{}{\Du{xxx}} , 
\label{Cextra'}\\
&& 0= \Dfactor{}{\Du{xx}} -\D{x}\Dfactor{}{\Du{xxx}} . 
\label{Cextra''}
\endEQs
Note that \Eqref{Cextra'} is a differential consequence of \Eqref{Cextra''},
while \Eqref{Cextra} 
combined with a differential consequence of \Eqref{Cextra'}
reduces to 
\EQ\label{Cextramore}
0=\Dfactor{}{u} . 
\endEQ
Consider \Eqref{Cextra''}.
Its highest-order term is $\Dfactor{}{\Du{xxx}\Du{xxx}} \Du{xxxx}$,
and hence $\Dfactor{}{\Du{xxx}\Du{xxx}}=0$. 
This yields, using \Eqref{Cextramore},
\EQ\label{Cfactorform}
\factor{} = a(x,\Du{x},\Du{xx}) \Du{xxx} + b(x,\Du{x},\Du{xx}) . 
\endEQ
Then the remaining terms in \Eqref{Cextra''} give
\EQ\label{Cbetaeq}
b_{\Du{xx}} = a_x +a_{\Du{x}} \Du{xx} . 
\endEQ

We now turn to the symmetry determining equation \eqref{kgsymmsys}.
This becomes, taking into account \Eqrefs{Cdependence}{Cextramore},
\EQs
0= &&
\solD{t}\Dfactor{}{x} +\Du{xx} \solD{t}\Dfactor{}{\Du{x}}
+\Du{xxx} \solD{t}\Dfactor{}{\Du{xx}} 
+\Du{xxxx} \solD{t}\Dfactor{}{\Du{xxx}} 
\nonumber\\&&\fewquad
+\Dfactor{}{\Du{x}} \D{x}g
+\Dfactor{}{\Du{xx}} \Dx{2}g 
+\Dfactor{}{\Du{xxx}} \Dx{3}g
-\factor{} g' . 
\label{Csymm}
\endEQs 
We substitute \Eqref{Cfactorform} into \Eqref{Csymm} 
and separate it into highest derivative terms in descending order.
The terms of highest order involve $\Du{xxxx}$, which yield
\EQ\label{Calphaeq}
0=a_{\Du{x}} g +a_{\Du{xx}} \Du{x} g' . 
\endEQ                                               
Since $g$ depends only on $u$, while $a$ does not depend on $u$,
the terms in \Eqref{Calphaeq} can balance only in two ways:
either $g$ and $g'$ are proportional, 
or $g$ and $g'$ are linearly independent 
with vanishing coefficients. 
This leads to two cases to consider.

\proclaim{ Case (i): $g,g'$ linearly independent }
In this case $a_{\Du{x}}=a_{\Du{xx}}=0$, so $a=a(x)$.
Consequently from \Eqref{Cbetaeq}, 
we have $b_{\Du{xx}} = a_x$,
and hence
\EQ\label{Cibeta}
b= a'(x) \Du{xx} + c(x,\Du{x}) . 
\endEQ
To proceed, we return to the symmetry determining equation \eqref{Csymm}.
The highest order remaining terms now involve $\Du{xx}$, which lead to
\EQ\label{Cigammaeq}
0=a' g' + 3a \Du{x} g'' + c_{\Du{x}\Du{x}} g . 
\endEQ
By taking a partial derivative with respect to $\Du{x}$
we obtain $0=3a g''+  c_{\Du{x}\Du{x}\Du{x}} g$.
Since $g$ depends only on $u$, while $a$ does not depend on $u$,
this equation immediately yields
\EQs
&& g''= \sigma g, \sigma=\const, 
\label{Cigeq}\\
&& c_{\Du{x}\Du{x}\Du{x}} = - 3a \sigma, 
\label{Cigammaeq'}
\endEQs
where $\sigma\neq 0$ in order for $g(u)$ to be nonlinear.
From \Eqref{Cigammaeq'} we obtain
\EQ\label{Cigamma}
c = 
-\frac{1}{2} \sigma a(x) \Du{x}{}^3 + c_2(x) \Du{x}{}^2
+c_1(x) \Du{x} +c_0(x) . 
\endEQ
Now we find that \Eqref{Cigammaeq} reduces to 
$0=a' g' + 2c_2 g$,
which immediately separates into the equations 
\EQ\label{Cialphagammaeq}
a' =c_2 =0 . 
\endEQ
Finally, the terms remaining in 
the symmetry determining equation \eqref{Csymm}
simplify to 
$0=c_{1x} g - c_0 g'$,
and thus we have
\EQ\label{Cigammaeqs}
c_0 = c_{1x} =0 . 
\endEQ

From \Eqref{Cfactorform}, \Eqref{Cibeta}, and \Eqsref{Cigamma}{Cigammaeqs},
we obtain
\EQ\label{Cifactor}
\factor{} = a (\Du{xxx} - \frac{1}{2} \sigma \Du{x}{}^3) + c_1 \Du{x},
a =\const, c_1=\const. 
\endEQ
This multiplier is admitted for all nonlinear interactions $g(u)$
satisfying \Eqref{Cigeq}
with $g(u)$ and $g'(u)$ being linearly independent. 
The general solution for $g(u)$ breaks into two forms:
\EQs
&& g(u)= \alpha\sin( \sqrt{|\sigma|} u + \beta), 
\eqtext{ if $\sigma<0$; }
\label{Cigsin}\\
&& g(u) = \alpha(e^{\sqrt{\sigma} u+\beta} \pm e^{-\sqrt{\sigma} u-\beta}), 
\eqtext{ if $\sigma>0$; }
\label{Cigsinh}
\endEQs
with arbitrary constants $\alpha,\beta,\sigma\neq 0$. 
By a scaling and a translation $\sqrt{|\sigma|}u +\beta\rightarrow u$,
and a scaling $\alpha\sqrt{|\sigma|} t\rightarrow t$,
we see that the resulting Klein-Gordon equation \eqref{kgeq} becomes,
respectively,
the sine-Gordon equation \eqref{sgeq}
or the sinh-Gordon equation \eqref{shgeq}. 
This completes the classification in case (i).

\proclaim{ Case (ii): $g,g'$ proportional }
In this case we have
\EQ\label{Ciigeq}
g'= \sigma g, \sigma=\const
\endEQ
where $\sigma\neq 0$ in order for $g(u)$ to be nonlinear.
Now \Eqref{Calphaeq} becomes
\EQ\label{Ciialphaeq}
0=a_{\Du{x}} + \sigma a_{\Du{xx}} \Du{x} . 
\endEQ
This is a linear first-order PDE for $a$ in $\Du{x},\Du{xx}$.
The solution is 
\EQ\label{Ciialpha}
a = a(x,\xi), \quad 
\xi=\Du{xx} - \frac{1}{2} \sigma \Du{x}{}^2 . 
\endEQ
Then from \Eqref{Cbetaeq} we obtain
\EQ\label{Ciibeta}
b= {\tilde a}_x(x,\xi) 
+\sigma\Du{x} ({\tilde a}(x,\xi)-a(x,\xi)\Du{xx}) + c(x,\Du{x}), \quad
{\tilde a} = \int a d\xi . 
\endEQ

To proceed, we return to the symmetry determining equation \eqref{Csymm}.
We find that the highest order terms involving
$\Du{xxx}{}^2$ and $\Du{xxx}$
vanish as a consequence of \Eqref{Ciialphaeq}. 
Then we find that the next highest order terms which involve $\Du{xx}$ 
in \Eqref{Csymm} yield 
$c_{\Du{x}\Du{x}}=0$. 
Hence, we obtain 
\EQ\label{Ciigamma}
c= c_1(x) \Du{x} + c_0(x) . 
\endEQ
Finally, the remaining terms in \Eqref{Csymm} reduce to 
\EQ\label{Ciigammaeqs}
0= c'_1 - \sigma c_0 . 
\endEQ

Thus, from \Eqref{Cfactorform}, \Eqrefs{Ciialpha}{Ciibeta}, 
\Eqrefs{Ciigamma}{Ciigammaeqs}, 
we obtain
\EQ\label{Ciifactor} 
\factor{} = 
{\tilde a}_\xi \Du{xxx} +\sigma \Du{x}( {\tilde a}-\Du{xx} {\tilde a}_\xi )
+{\tilde a}_x 
+c_1 \Du{x} + \frac{1}{\sigma} c'_1, \quad
\xi=\Du{xx} - \frac{1}{2} \sigma \Du{x}{}^2 , 
\endEQ
with arbitrary functions $c_1(x),{\tilde a}(x,\xi)$. 
This multiplier is admitted for all nonlinear interactions $g(u)$
satisfying \Eqref{Ciigeq}.
The general solution of this equation is 
\EQ
g(u) = e^{\sigma u+\beta}
\endEQ
with arbitrary constants $\beta,\sigma \neq 0$. 
By a scaling and a translation $\sigma u+\beta\rightarrow u$, 
and a scaling $\sigma t\rightarrow t$,
we see that the resulting Klein-Gordon equation \eqref{kgeq} becomes 
the Liouville equation. 
This completes the classification in case (ii).
\endProof

\begin{acknowledgments}
The authors are supported in part by 
the Natural Sciences and Engineering Research Council of Canada.
We gratefully thank the referees for useful comments
which have improved this paper. 
\end{acknowledgments}

\end{document}